# Pegadas: A Portal for Management and Activities Planning with Games and Environments for Education in Health

T. K. L. COSTA, Federal University of Paraiba, Department of Exact Sciences
L. S. MACHADO, Federal University of Paraiba, Department of Informatics
A. M. G. VALENÇA, Federal University of Paraiba, Department of Clinical and Social Dentistry
M. A. WINCKLER, Université Nice Sophia Antipolis
R. M. MORAES, Federal University of Paraiba, Department of Statistics

Applications for learning and training have been developed and highlighted as important tools in health education. Despite the several approaches and initiatives, these tools have not been used in an integrated way. The specific skills approached by each application, the absence of a consensus about how to integrate them in the curricula, and the necessity of evaluation tools that standardize their utilization are the main difficulties. Considering these issues, Portal of Games and Environments Management for Designing Activities in Health (Pegadas) was designed and developed as a web portal that offers the services of organizing and sequencing serious games and virtual environments and evaluating the performance of the user in these activities. This article presents the structure of Pegadas, including the proposal of an evaluation model based on learning objectives. The results indicate its potential to collaborate with human resources training from the proposal of the sequencing, allowing a linked composition of activities and providing the reinforcement or complement of tasks and contents in a progressive scale with planned educational objective-based evaluation. These results can contribute to expand the discussions about ways to integrate the use of these applications in health curricula.

CCS Concepts: • **Applied computing** → **Learning management systems**; *Interactive learning environments*; • **Software and its engineering** → **Abstraction, modeling and modularity**; Interactive games; Use cases;

Additional Key Words and Phrases: Technology to support education and training, serious games, educational games, virtual environments, portal of games and virtual environments

# 1 INTRODUCTION

Serious games (SG) and virtual environments (VE) can be used as a way to teach and train skills as a complement to the traditional ways of learning. Challenges related to the use of SG and VE are presented by Nunes et al. (2014) and Graafland et al. (2012) and include their integration in curricula, their validation, and the automatic evaluation of students. Previous research has presented the advantages of using SG and VE for the teaching and learning processes, showing their effectiveness in learning (Graafland et al. 2012; Bellotti et al. 2013; Nunes et al. 2014). However, Hays (2005) points out, from research on one specific case, that we should not generalize effective learning by means of games to all situations and for every instructional task. It is known that those applications are developed to consider a specific problem or subject, and it usually focuses on particular content of a discipline. Additionally, the level of the approach can depend on the stage of the course for which the application was designed.

Different SG and VE have been built and tested as resources to support the process of teaching and learning in health, especially in the medical field (Machado et al. 2011). Through SG and VE, students and professionals can train skills related to the cognitive, affective, and psychomotor domains (Arnab et al. 2013). This approach enables experimental praxis, reducing the distance between theory and practice, since it allows students to learn by simulated activities (Arnab et al. 2013; Pfahl et al. 2000). The interest in patient safety and satisfaction is another factor that motivates the use of applications for education in health (Satava et al. 2003). In this context, by the use of SG and VE, the early identification of errors in simulated environments (rather than in real patients) is possible, increasing students' and practitioners' confidence in dealing with real-life cases. In the traditional context of learning and health qualification, evaluating students under the same conditions in a standardized way is difficult, and it is observed that procedures are evaluated based on the subjective and particular interpretation of the professor (Cosman et al. 2002). Thus, SG and VE can provide an alternative for performance evaluation based on objective measures (Moraes and Machado 2013).

Currently, the access to SG and VE on the Internet is done through web portals or platforms designed for collection and distribution. These portals and platforms allow the search of SG and VE, but they still do not provide means to manage different types of resources, such as planning, organization, and evaluation of activities involving the integrated use of these applications. Particularly for the health area, some of these resources are available from company websites to promote their own applications. The Games and Simulation for Healthcare Library and Database (http://healthcaregames.wisc.edu/index.php) is an example of a website that provides links to SG and VE developed by different companies and research groups. In this website, it is possible to find the applications and download and use them in an independent way. In a broader sense of education, it is possible to find portals and platforms for SG content customization, such as Pingo (Nogueira et al. 2013); storage, search, and statistics of the most played games by users, such as Attractive Virtual Educational Portal (Levashenko et al. 2013); and game development and sharing environments, such as Simurena (Wagner 2012). However, it was observed that none of them allows for the planning of sets of activities, organization of the contents in an ascending order of difficulty, and evaluation of users considering the utilization of the applications in an integrated way.

This way, taking advantage of the benefits of SG and VE for education and training in health, and the potential for expanding the functionalities of the current tools that group SG and VE, this article presents Portal of Games and Environments Management for the Design of Activities in Health (Pegadas), an educational portal for the grouping of SG and VE for health sciences. It was developed to assist in the planning, management, and evaluation of users in sequences of activities for training by SG or VE. This evaluation is based on educational objectives classified in the different domains of learning: cognitive, affective, and psychomotor.

## 2 METHODOLOGY

The search for ways to assist in the planning and monitoring of activities that include the integrated use of SG and educational VE is an emerging area that generates different outcomes in the search for solutions. The design of Pegadas follows an applied research style called *presentation of something different* by Wazlawick (2009).

In its design process, a few basic steps were followed to plan, produce, and test Pegadas. At the initial phase, the educational objective for health was considered the main focus of the portal. The development was divided into five stages as follows: (1) the definition of educational skills that can be developed by the use of SG and VE for health; (2) definition of an evaluation method; (3) definition of the structure and the main modules of Pegadas; (4) implementation; and (5) tests.

The first stage considered Miller's pyramid of clinical competences (Miller 1990) and included brainstorming about the skills that are necessary in the educational process in health. The second stage consisted of the definition of a method of evaluation able to consider these skills in the evaluation process. The structure of Pegadas was defined in the third stage, which included the design of a set of modules for management and organization of tasks and users' evaluation in a tool that supports the integrated use of SG and VE.

The implementation (fourth stage) included the selection of tools for the portal's development. HTML[1] and CSS[2] were used for web interface; PHP[3] for web communication with the server; JavaScript,[4] JQuery,[5] and AJAX[6] for dynamic resource deployments in the portal; and MySQL for database management. These tools were selected due to their capability in providing more flexibility in the development of web applications and to guarantee the functioning of the portal in different operational systems. In the fifth stage, tests were conducted to guarantee the functionality of the pages, functionality of each module (module test), functionality of the portal (integration test), and compatibility with different browsers (compatibility test).

## 3 DEFINITION OF EDUCATIONAL SKILLS THAT CAN BE DEVELOPED BY THE USE OF SG AND VE FOR HEALTH

The process of professional education in health requires the training of different types of situations that may be experienced through the course and in practice (Eraut 1994; Amâncio Filho 2004; Santos 2011). It is necessary to consider the importance of preparing a professional that is capable of adapting to the constant changes in the area. This fact contributes to the understanding that the educational process, from the beginning, needs to include practice in real or simulated contexts (Fernandes et al. 2003).

To support the educational process, SG and VE can simulate contexts in which the application of knowledge is necessary for the execution. They can also be used to simulate different competences, improve skills (as psychomotor abilities), enhance the process of decision making (those that demand fast response), reproduce unusual situations (simulation of rare cases), and so on (Machado et al. 2011).

Miller's pyramid (1990) is a way to classify clinical competence in educational settings or in the workplace. The structure of the pyramid distinguishes knowledge and action in levels (Figure 1). It presents "knowledge" as a foundation for the process of competence acquisition, which involves obtaining the necessary knowledge to perform tasks efficiently. The second level is aligned with the ability to know how to use knowledge, that is, contextualized knowledge ("knows how"). The third level refers to behavior ("shows how"), that is, to expose how to act in front of the presented problems. In the last level, there is the "does," which is demonstrated in real situations from the usual practice (Miller 1990). Miller's ideas strove to define education by outputs and not by inputs. At the end of any teaching intervention, the interest is in what learners can do. The highest levels are related to highest professional authenticity.

---

[1] *HyperText Markup Language*: http://www.w3schools.com/html/html_intro.asp.
[2] Cascading Style Sheets: https://www.w3schools.com/css/.
[3] *PHP: Hypertext Preprocessor*: http://php.net/.
[4] https://developer.mozilla.org/en-US/docs/Web/JavaScript/Guide/Introduction.
[5] https://jquery.com/.
[6] *Asynchronous Javascript and XML*. Técnica: http://www.w3schools.com/xml/ajax_intro.asp.

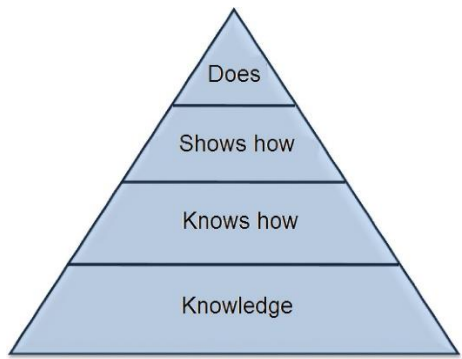

Fig. 1. Miller's pyramid.

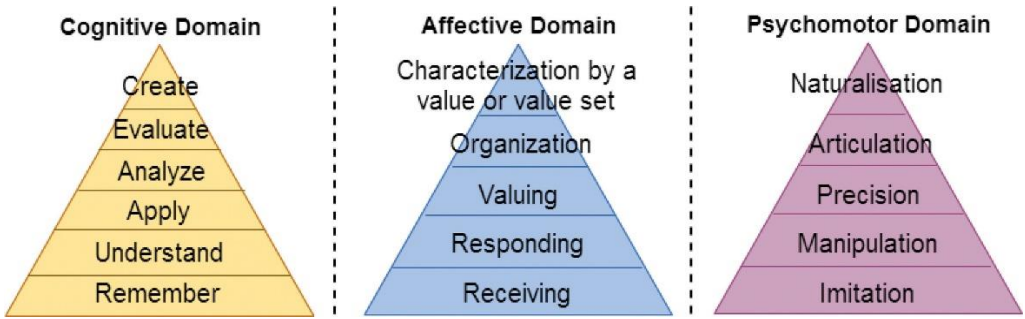

Fig. 2. Domains of learning and categories.

Miller's pyramid can be seen similarly to Bloom's taxonomy, as it may help to understand what is being tested and whether the evaluation method used is valid for the assumptions. The hierarchy of learning outcomes, which appears in Miller's pyramid, had already been discussed as a form of classification in Bloom's Taxonomy (Bloom 1984). According to Bloom, the integral formation of an individual involves three basic domains: cognitive (knowledge), psychomotor (skill), and affective (attitude). These domains can be experienced simultaneously, helping to acquire competences. Bloom's Taxonomy is widely used in the literature, allowing the exposure of the objectives by different people to be clearly standardized, revealing their importance in the context of learning domains, and helping in the elaboration of the evaluation tool. The domains of learning are divided into categories, as shown in Figure 2.

The cognitive domain involves learning and includes obtaining new knowledge, intellectual ability, understanding, and thinking about an issue or fact. It was revised by a group of experts supervised by Krathwohl, one of the original authors. The revised taxonomy took the basic idea of the cognitive domain and divided it into two dimensions: cognitive process dimension and knowledge dimension. However, the cognitive process dimension is better known and applied in academia. This dimension is related to the process, in other words, it is related to "how" something is achieved, representing incremental cognitive complexity. It is divided into six hierarchical categories: remember, understand, apply, analyze, evaluate, and create (Ferraz and Belhot 2010; Krathwohl 2002).

The affective domain is related to feelings, attitudes, and values, that is, the manner in which we deal with things emotionally and culturally. It can be demonstrated by behaviors and attitudes of awareness, interest, attention, concern, responsibility, ability to listen and respond to interactions with others, and the ability to demonstrate those values appropriately to each type of test and field of study (Bloom 1984; Krathwohl et al. 2002). The affective domain is divided into five hierarchical categories: receiving, responding, valuing, organization, and characterization by value or value set.

In turn, the psychomotor domain involves skills that combine muscular actions, dealing with abilities of handling tools or objects. Bloom and his colleagues have never created subcategories for skills in the psychomotor domain, but other educators contributed to this third domain. An example is the psychomotor domain proposed by Dave (1970), which is simple and suitable for most adult training in health. It is divided into five categories, which are imitation, manipulation, precision, articulation, and naturalization.

The use of the Taxonomy of Educational Objectives (also denominated Bloom's Taxonomy) in Pegadas does not delimit any educational modality or strategy. Thus, the concern is restricted to the effectiveness of the process. In this way, this taxonomy demonstrates its relationship with the "how" to apply and evaluate objectives instead of restricting forms or environments in which learning can occur (Ferraz and Belhot 2010). SG and VE have a set of objectives that can be classified based on Bloom's taxonomy, helping mainly in two aspects: selection and evaluation.

## 4 EVALUATION METHOD

In Pegadas, there is a sequencing of activities that consists in organizing tasks and/or planned contents according to their educational objectives. The activities that make up the sequence correspond to the selected SG and VE, which can be gathered and organized into levels of complexity (Costa et al. 2016a).

The levels have a hierarchical organization with each one consisting of a set of SG and VE. The progress in each level is possible only after the accomplishment of the minimum requirements. In the same level, the SG and VE gathered have no restrictions in the order of execution. Thus, students can choose and execute the activities of the same level in the order they want.

The mediators are responsible for defining the minimum requirements for the learner to pass to a next level, and these requirements are represented by minimum values of performance. To achieve that, the learning domains were used to define categories that can be selected by the mediator. At the conclusion of the sequence, students will demonstrate that they have achieved the minimum requirements and purpose of the activity sequence.

Figure 3 shows an example of a sequence of activities composed by three levels. The first level consists of four activities (SG or VE); the second level is composed of three activities, and the last level of two activities. The SG and VE that appear in the first level can be executed according to the student's preferences. However, performing the second level's activities is only possible when the student achieves the minimum requirements in the first level (specified by the mediator). The process is repeated at the other levels of the sequence, allowing the mediator to plan activities that require different depths of knowledge and complexity. These activity sequences can be based on a content extracted from one of the axes to be worked in the process of teaching and learning in health, appropriate to the scope and planned by the mediator.

The evaluation of the activity sequence is proposed to be performed automatically and must be personalized by the mediator. The analysis is done at each level and based on the achievement of the educational objectives present in the individual activities and that the mediator selected to be considered during the evaluation process. Thus, when an activity is completed, the system can (re)calculate the learner's performance in the current level from the success or failure in the SG/VE (Figure 4). The educational objectives declared by different SG and VE were classified according to Bloom's Taxonomy (Bloom 1984; Krathwohl et al. 1984; Krathwohl 2002).

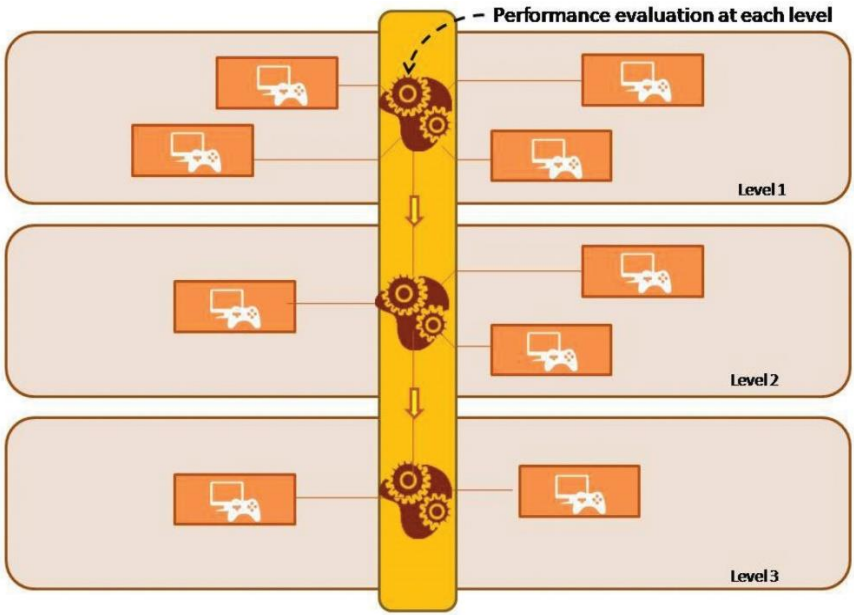

Fig. 3. Activities sequence example.

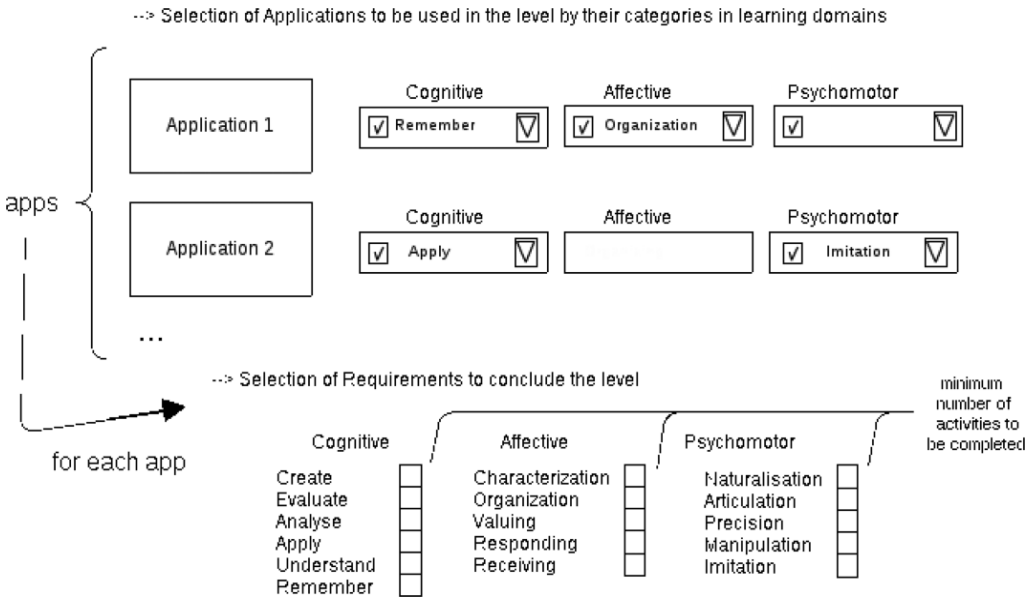

Fig. 4. Example of how applications can be used according to the learning objectives in each domain.

The process for analysis and decision-making in the evaluation module is carried out by a Rule-Based System (RBS). The RBS consists of a set of statements that form a "working memory" and a set of rules that guide the actions according to the statements. The strategy used is the knowledge codified in a set of rules (Ligeza 2006).

Generally, these systems are simple models that can be adapted for different cases, especially for cases in which all the knowledge in the area can be described through rules. According to Millington and Funge (2009), the creation of an RBS should consider a relevant set of facts, based on knowledge, and it is represented by rules, which cover every possible action. Rules are composed by conditions that must be analyzed to guide particular actions.

In the execution of the behavior represented by the rules, each rule is executed to verify an activity presented in a level. This way, the evaluation module performs the analysis at each level and then verifies whether the rules' conditions have been met. If positive, then the player moves to the next level; if negative, then the player remains at the same level.

The evaluation module checks the student's performance at each level of the sequence of activities. Performance is measured based on the achievement of educational objectives established for each level. The definition of objectives is made by the mediator in two steps:

1. Selection of categories of learning domains they want to evaluate in each game or environment. These categories represent the educational objectives that the mediator wishes to work with their mediated;
2. Determination of degree of minimum sufficiency in each educational objective established. This degree of sufficiency is represented by a minimum success value that a student must obtain in each category that will be evaluated (relevance of the objective in planning).

The rules will be applied based on the attributes established by the mediator. Therefore, for each category selected in the evaluation:

- There will be the calculation of the player's success in the category;
- Verify that the rule was complied with.

The general rule for each category is as follows (Costa et al. 2016b):

---
**ALGORITHM 1:** General Rule for Each Category
---
**if** success_value_of_user_in_category **is greater than or equal to** minimum_success_value_defined_by_the_mediator_for_the_category
      **then** the_category_objective_was_achieved
      **else** the_category_objective_wasn't_achieved
**end**

After analyzing all categories, a new rule is verified as follows:

---
**ALGORITHM 2:** General Rule after Analysis of all Categories
---
**if** all_objectives_achieved
      **then** level_complete
      **else** level_incomplete
**end**

# 5 THE STRUCTURE OF PEGADAS AND ITS MAIN MODULES

Pegadas is based on components and properties for educational support software with performance evaluation in activity sequences that involve the use of SG and VE for education and training. This portal intends to help mediators (i) integrate serious games and virtual environments in the planning of their didactic actions and (ii) monitor the performance of users during the activities. The choice of name is due to the fact that this instrument provides planning of a learning trail. When walking along a learning trail, presented by a mediator, the student performs activities that help their training process and leaves his or her "footprints" (*pegadas* in Portuguese).

A conceptual map was produced (Figure 5) to represent the set of ideas and concepts in Pegadas, arranged into a kind of network of propositions. This conceptual map shows that the portal manages the activities, trails (also denominated in this work as activity sequences), users, relationships, and information generated by the users. Each activity can be a serious game or a virtual environment, and it has well-defined educational objectives that follow the classification according to Bloom's Taxonomy. The trails, in turn, have levels composed of activities and have an evaluation model based on the achievement of educational objectives. The users of this portal are the mediators, students, and developers with access to specific services.

Pegadas' general structure consists of six modules, shown in Figure 6: evaluation, database management, administrator, developer, mediator, and student (Costa et al. 2016a). The database management module contains the system's managers and databases, which are divided into the following: user manager, game/environment manager, activity manager, bank of users, and bank of games/environments. The user manager is responsible for controlling the user bank, which stores information of developers, mediators, and students. This managing submodule controls the creation, modification, and removal of records of users. The SG and VE manager is responsible for controlling the database that stores these resources (bank of games and environments). This managing submodule controls the registration, modification, and removal of available applications. To control the relationships among resources and users, there is the activity manager. This manager will receive requests from other modules and manage the communication with the user manager and the game/environment manager.

The developer module is responsible for the request of registration of SG and VE and, for this, it will communicate with the database management module and the administration module. In addition, it will allow the developer to access information from registered games and environments. The administration module is related to the developer and database management modules, which are responsible for the examination of registrations for approval of new applications.

The mediator module controls the necessary actions for the creation of sequences of activities, as well as the administration of students and classes, communicating with the manager and evaluation modules. The student module will allow students to access activity sequences, enabling their realization. This module will also allow access to games and environments individually through calls to the administration module. In addition, through this module, students who lack guidance will be able to request help from a mediator.

Finally, the evaluation module is responsible for customizing attributes of evaluation for each sequence of activities, acting, in this context, together with the mediator module. In addition, the evaluation module will also conduct a performance analysis of the student during the sequence, requiring a joint action with the student module.

## 5.1 Users and Personal Information Management

In Pegadas, there are three main types of users: mediator, student, and developer of serious games or virtual environments (Costa et al. 2016a). Each one of these profiles has different rights and duties in the system. Thus, according to the type of user, a specific group of services is accessible.

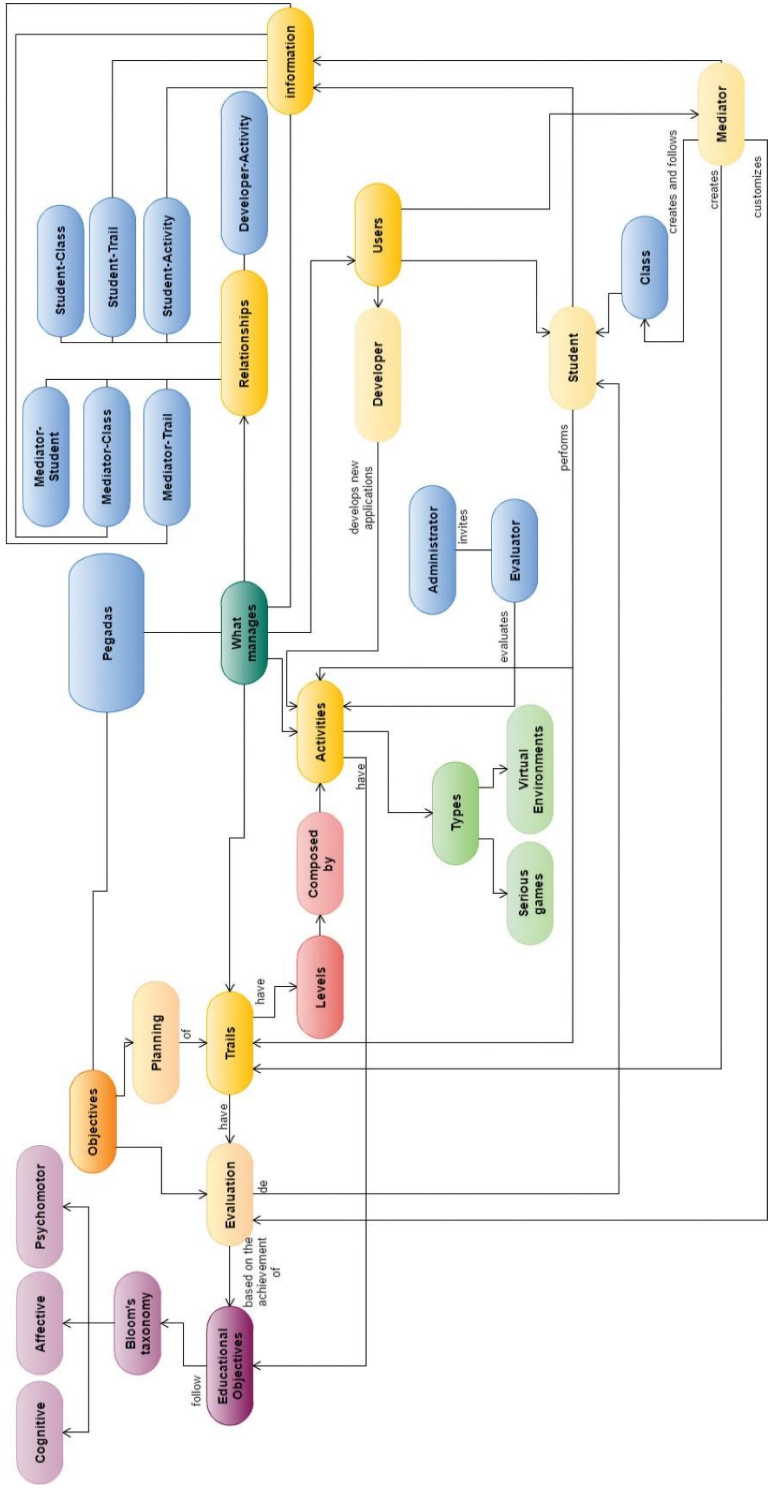

Fig. 5. Conceptual map.

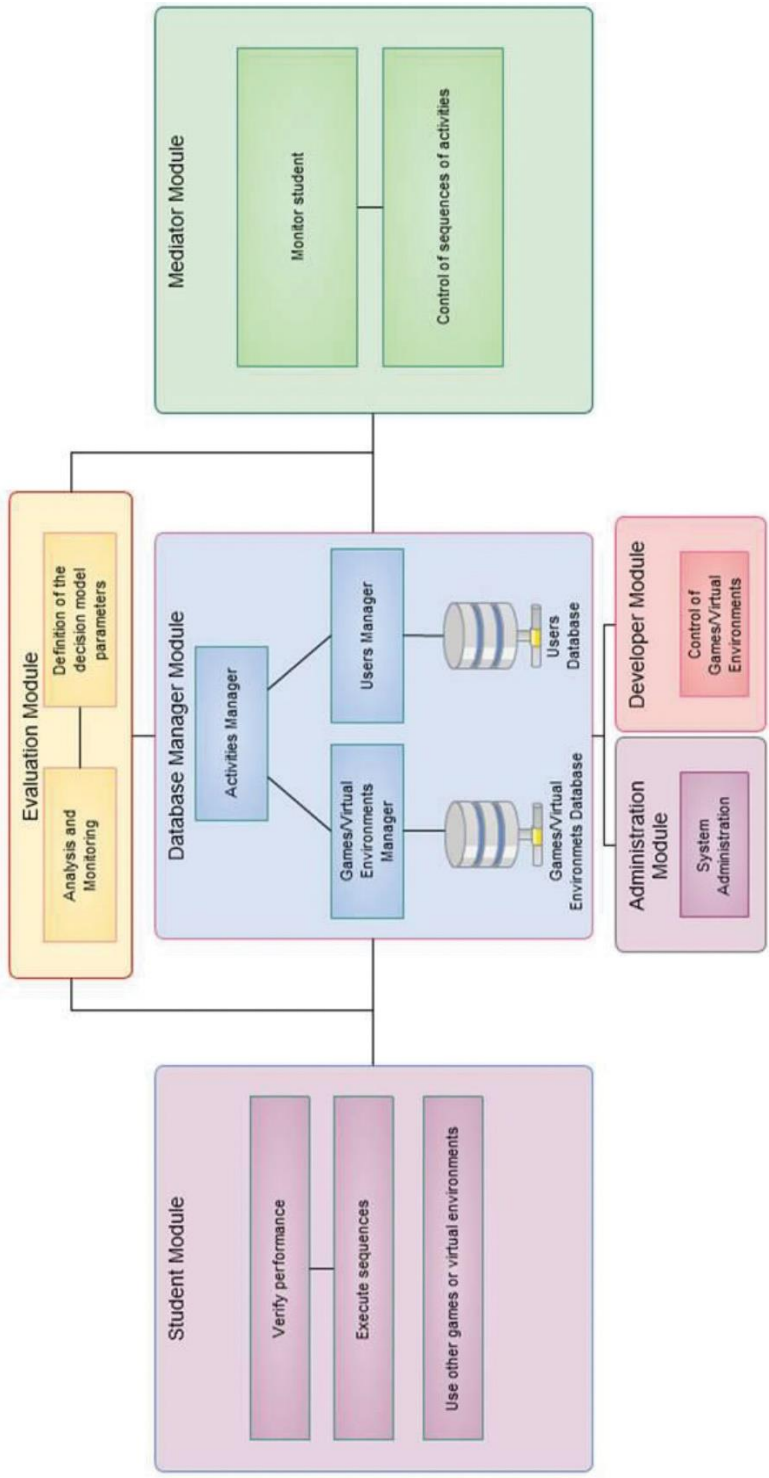

Fig. 6. Structure of Pegadas with its modules.

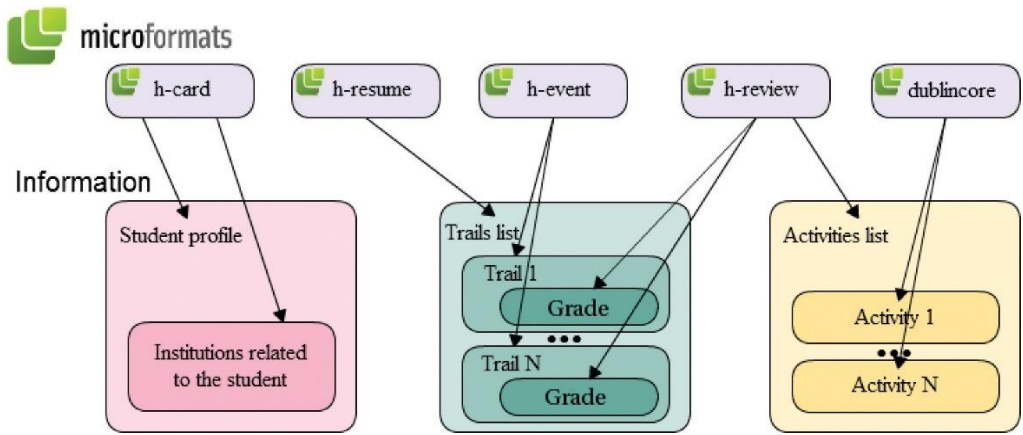

Fig. 7. Personal information and microformats used in Pegadas.

The mediator is the profile of the educator responsible for orienting students. This way, this user profile is able to manage students and search for available SG and VE and sequences of activities (in the form of serious games and virtual environments), adapting performance evaluation and monitoring students' performance. Students, in turn, can perform the sequences of activities proposed by the mediators, as well as independently search and select another SG and VE. Developers are able to request the insertion of new SG or VE and may also view or remove the applications that they have already inserted. The registration request of new tools requires the approval of the system administrators, who verify the suitability of the application to the context of Pegadas.

The SG and VE accepted in the system are available to be accessed by mediators and students. It is worth emphasizing that only the mediators can create sequences of activities, as well as adapt the evaluation model that analyzes the performance of the student along the sequence. However, both profiles (mediator and student) can perform searches, view details, and use the applications individually.

The personal data provided by the user and their accessibility contribute to the creation of collections of personal information. Students, for example, need to manage their personal collections of information in the various digital environments they use, and these digital environments must provide a means to facilitate the management.

Regarding the context of personal information management, Pegadas offers the service of interoperability of information between systems. The interoperability allows for the communication between the portal and a personal information management system, requiring the standardization of information.

To assist in information management and information compatibility among different systems, Pegadas standardizes some information. The standardization proposal was based on microformats (Prabhakar 2005) and it is supported by other learning aid environments that also use microformats as standard for data (Ermalai et al. 2013; Dragulescu et al. 2011; Tomberg and Laanpere 2009; Ermalai et al. 2009). In the context of Pegadas, the microformats suggested as standard for the information are as follows (Figure 7):

- h-card: simple format for publishing people and organizations on the web. Pegadas uses it to standardize the information (i) of the user profile and (ii) of educational institutions.
- h-resume: format for publishing résumés and résumés' summaries. Pegadas uses h-resume as a list of performed/created activity trails.

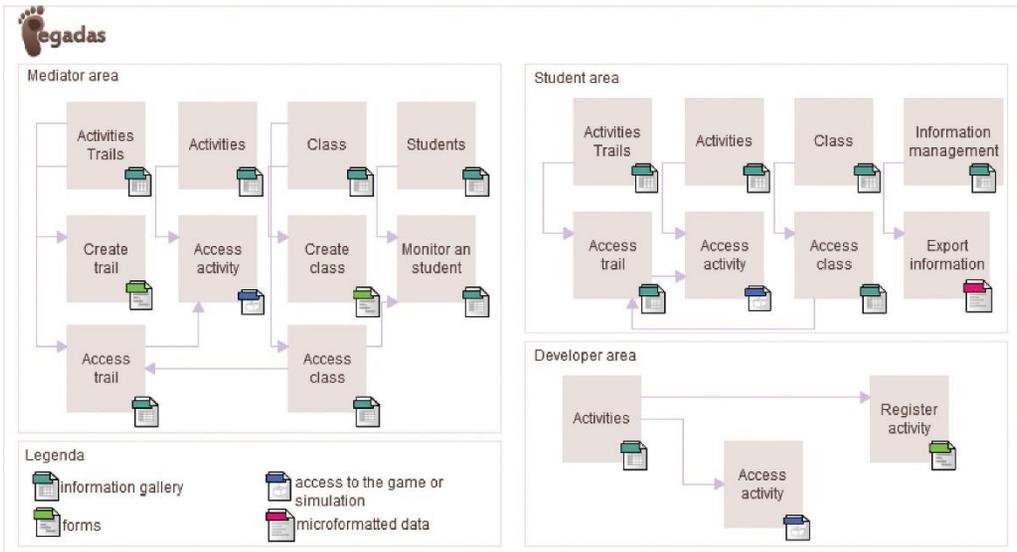

Fig. 8. Different areas of work: mediator (teacher or tutor), student, and developer of SG/AV.

- h-event: format for publishing events on the web. Pegadas uses it to describe every trail present in the list, as it is nested to the h-resume microformat.
- h-review: format for publishing reviews on the web. In Pegadas, h-review standardizes the information about (i) the student's grade in an activity trail, being, in this case, nested to the h-event microformat; and (ii) the list of activities.
- dublincore: format to describe resources. Pegadas uses it to standardize the information of the activities, that is, of the serious games and virtual environments. In this case, this format is nested to h-review when used to describe activity lists.

These standards help to organize information related to trails, activities, and report grades. By adopting microformats, users can export their data to other systems, promoting the interoperability of information.

## 6 IMPLEMENTATION

Pegadas was developed based on task models and portal screen prototypes. The task models aimed to highlight the services offered for each user profile. This model was elaborated through a task-modeling tool called Human-centered Assessment and Modeling to Support Task Engineering for Resilient Systems (HAMSTERS).[7] The screen prototypes aimed to define the layout of screens for each portal user's profile. Those prototypes were developed with the aid of the prototyping tool called Balsamiq Mockups.[8]

The three profiles (mediator, student, and game developer) represent user roles. Each of these profiles controls what the user can perform in the system, so each user must perform an authentication in the portal through login and password. After the authentication, the pages are displayed according to the profile. Due to security issues, users can only access pages related to their profiles. Figure 8 shows a diagram of the different pages of Pegadas and which profile can access them.

---

[7] https://www.irit.fr/recherches/ICS/softwares/hamsters/.
[8] https://balsamiq.com/products/mockups/.

Mediators are the only ones able to create activity trails; however, students can access the list of activities registered in the system to perform any activity of particular interest. At first, any user can register in the system, but only professors/teachers can create classes. Developers, when registering an activity, receive the guidelines on how to properly include the SG/VE in the portal, enabling communication among them.

### 6.1 Developer Module and Communication with SG and VE

The developer module allows the professional or team that develops health SG/VE to request the registration of their applications, to consult their list of registered activities, as well as to check the status of moderation by those in charge of Pegadas. To register the applications, the developer needs to provide information, such as activity name, description, area, type (whether it is a SG or a VE), format, language, rights/license observations, educational objectives, and a representative image of the activity. This information is stored in the database and follows the dublincore microformat standard, used to describe digital materials, since it has a simple and robust set of attributes that helps in the moment of retrieving some information or when cataloging these objects in a database (Ferlin et al. 2010).

During the registration, a compressed file containing the serious games or virtual environment must also be uploaded. Currently, the portal accepts SG and VE in the Unity Web Player or Construct 2 formats. For the adaptation to the portal, each application needs to send a variable indicating whether the player has succeeded or not in the activity. Through this variable, the portal evaluation model verifies whether the educational objectives of the SG or VE were reached to, subsequently, proceed with the analysis of the student's performance in the activity trail. At the time of the registration, the developer receives the necessary guidelines to adopt this pattern of communication with Pegadas.

### 6.2 Mediator Module

The mediator module allows the professor (or tutor) to register classes and activity trails (also called sequences of activities), in addition to having access to services, such as management of classes and trails created by him/her, as well as the visualization of students and activities. To create classes, they must enter the name, description, and the students who are associated with them. There is no restriction on whether a student is associated with more than one class. Classes can have data changed by the mediator or can be removed if the class has no trails in progress. Removing a class does not remove associated students from the portal.

The creation of activity trails is divided into two steps. In the first step, the mediator only names and describes the trail. In step two, the mediator selects the activities for each level and defines the evaluation for each level.

To define the evaluation (still in step two), the mediator must follow two steps at each level based on the proposed performance analysis presented in the structure. The mediator must

(1) select which categories of learning domains will be considered in each SG/VE and (2) define minimum success values in the selected categories.

In the first stage of step (2), the categories of educational objectives of interest for the evaluation are selected. Then (second stage of step (2)), there is the definition of the minimum number of activities that need to be performed as objectives for the level.

In addition to creating the activity sequences, the mediators can view the system's SG and VE. The list contains the name, an indicative image, description, area, and educational objectives. In the "Learn more" option, a page with all the SG or VE data are displayed, in which they are able to try the activity to better understand the gameplay and how educational content is shown to players.

## 6.3 Student Module

The student module allows students to access and perform the activity trails available to their class. The list of trails has information such as groups linked to the trails, number of levels of the trail, the student's level in the trail, completion deadline, and grade (if granted by the professor). In addition, they can visualize the description and educational objectives contemplated by the activities of the trail.

When accessing a particular trail, the student checks the levels as well as the activities of each level. They can also observe the completed levels, what is in progress, and the following levels not yet activated for performance. When starting a trail, the student visualizes the progress of the first level, but the other levels remain inactive until the student completes the first level.

When viewing the trail, the student can access a particular activity and check all the data from that SG or VE. Choosing to perform the activity, the student is directed to the page in which the SG or VE is encapsulated, being able to execute it. After its completion, the result is used by the evaluation module that updates the activity trail, as shown in Figure 9.

In relation to the information management service, the student verifies his/her profile data, trails, and activities (Figure 10). This information is standardized through the microformats defined in the structure and can be downloaded by the student for use in a personal information management system or in another system that accepts data with the same standard.

## 7 TESTS

An analysis of the evaluation method of the student was performed by a team of five experts. These experts analyzed the form of evaluation adopted by the system. For them, the choice of Bloom's Taxonomy as a means of classifying educational objectives and as a means of assisting the achievement of students' performance evaluation is an adequate alternative for monitoring students in a learning support environment. In addition, they verified the proper functioning of the system evaluation module during a test session.

The execution of additional tests was performed to check four main elements of the system: page appearance, module functions, page integration, and browser compatibility. The tests were performed with commodity computers in a conventional network (7Mbytes/s) with a single user connection. Since the tests were performed to verify the portal functioning and the fact that the pages and applications run locally in the user's computer, the network capabilities were not evaluated. The page appearance tests included the check of page orthography and links in buttons. This test allows us to correct grammar and orthographic errors, broken links, and bad positioning of graphical elements.

The module tests were performed in small parts during the pages' development. These tests considered the execution of the module task and checked their ability to answer queries in the web environment. In relation to the developer module, the main test was related to the registration of activities, in which the consistency of the process of storing the activities in Pegadas's database was verified. These tests were succeeded by tests that verified the compatibility of the registered activity with Pegadas's portal. Thus, the operation of SG and VE in Pegadas has been ratified, as long as they follow the standard communication with the portal (presented to the developer at the time of registration). The tests also contribute to the professor and student module, since both users have access to the system's activities and can execute them from Pegadas. In the mediator module, the main test was related to the creation of activity trails. This test verified the consistency of the data stored in the database to be used in later queries. For the student module, an important performed test was the collection of information on the users' performance in SG and VE, as well as their use in the evaluation module, responsible for updating the activity trail monitoring.

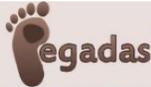

Fig. 9. Preview of an activity trail with the first level concluded and the second level in progress.

After module tests, page integration tests were performed on the system, especially checking for data transmission failures between the pages and the database. Finally, compatibility tests with three browsers (Mozilla Firefox, Google Chrome, and Internet Explorer) were performed. The browsers that are currently best suited for viewing the portal are Firefox and Chrome. Internet Explorer still presents instabilities that need to be repaired.

The tests prove that it is possible to evaluate the student's performance in sequences of tasks and content presented in the form of serious games and virtual environments. This shows the effective accomplishment of offering the possibility of evaluating users based on educational objectives from the integrated use of those resources.

## 8  DISCUSSION AND FINAL REMARKS

Different research fields have emerged with the use of serious games and virtual environments to support education and training, particularly in medicine. Among them, there is research on ways to assist the planning and monitoring of activities that include the integrated use of SG and VE for educational support. In this field, the present article has presented Pegadas, a web portal that allows the sequencing of tasks and contents in the form of SG and VE and is able to evaluate the performance of users in their learning process. Among the contributions, the article highlights:

- Discussion about educational skills in different domains that can be developed by the use of SG and VE for health;
- Evaluation model, in the context of activity sequences, that promotes a performance analysis of the integrated use of SG and VE from different domains and skills (important for the medical field);
- Standardization of the collection of personal information, assisting in the management and interoperability of information among systems.

Fig. 10. Page of information management with option to download standardized data.

The evaluation model promotes the analysis of performance along the activity trails. This analysis allows the following:

- A comprehensive assessment that considers educational aspects of different serious games and virtual environments to analyze the scope of knowledge, skills and attitudes;
- Multiple possibilities for fulfilling the requirements, based on establishing the degree of sufficiency for the achievement of objectives.

Educational objectives should be present in both planning and evaluation of learning. Thus, as support activities for learning, the SG and VE with educational purposes in the medical field have contents and practices aimed at reaching educational objectives that involve more than the cognitive domain, passing through the psychomotor and affective domains. Thus, in Pegadas, the objectives are classified through the Taxonomy of Educational Objectives, which helps to differentiate domains of learning and levels of knowledge, skills, or attitudes aimed for health professionals. The evaluation model allows monitoring students' performance and automating the control of progress in the levels of an activity trail. For this, the model uses educational objectives as evaluation control requirements. The performance analysis considers the different activities present at each level of the sequence (i.e., a comprehensive evaluation). The customization that can be performed for the evaluation allows defining the degree of requirement to complete each level and, consequently, the sequence (or trail) of activities. Depending on the requirement set, the student can have multiple paths to achieve the objectives.

In relation to the grouping of SG and VE into a single system, the activity trails are presented as a way of bringing them together, proving to be useful in sequencing that allows planning and organization at levels that can be differentiated by objectives, skills, and complexity of content. The possibility of level planning allows for two important forms of organization. The first corresponds to the alternative of gathering activities in the same level, in which all SG and VE can be performed without prerequisites. In the second form of organization, the activities are leveled, that is, the groups with SG and VE for educational support must consider levels or prerequisites among them. The union of those forms of organization allows for grouping and leveling different sets of SG and VE, promoting flexible planning for different purposes. This way, this service enables the expansion of the scope of contents that can be explored and experienced, adding previously acquired competencies and encouraging the emergence of new knowledge, skills, and attitudes.

The third contribution highlighted in the article is related to the management of personal information. As students deal with a significant amount of personal information throughout their academic life, which are fragmented in the different environments they use, the work aims to standardize the main information acquired and generated by Pegadas. Standardization is used as an integration solution, aiding in the interoperability of information. Pegadas has intended to standardize the semantic interoperability by adopting microformats, since it is a standard already used in other learning support environments. The adoption of a standard already used by other environments and tools strengthens the interoperability and encourages adherence to new environments.

The planning of assistance services, through the creation of activity trails, and the service of performance analysis on the course of the trail, give Pegadas new characteristics. The definition of the portal scope as a tool to support medical education highlights the need to popularize (and further develop) SG and VE for this area. The creation of a database with this content would allow the implementation of new resources in the portal, such as access and search of resources by subject in the various areas of medicine. Although the results have demonstrated the effectiveness of the sequencing of activities with the performance analysis of its users, and the tests performed have proved the functionality of the main services, the implementation of these new features in the portal is necessary to perform tests with different user profiles.

The planning of assistance services, through the creation of activity tracks, and the service of performance analysis in the path of the trail, show innovative characteristics of Pegadas. Despite the previous research and contributions, the automatic evaluation of the integrated use of SG and VE had not yet been considered in health training. The proposal of a standardized method of evaluation and ways of combining applications developed for health learning can contribute to broadening the discussions on ways to integrate the use of training applications into health curricula.

Currently, there is only the Portuguese version of the portal, but, in the future, it may be available in other languages.